%% file: main.tex
\documentclass[sigconf, nonacm]{acmart}

\newcommand\vldbdoi{10.14778/3551793.3551834}
\newcommand\vldbpages{2826 - 2838}
\newcommand\vldbvolume{15}
\newcommand\vldbissue{11}
\newcommand\vldbyear{2022}
\newcommand\vldbauthors{\authors}
\newcommand\vldbtitle{\shorttitle} 
\newcommand\vldbavailabilityurl{https://github.com/Cyril-Tang/CRC-query}
\newcommand\vldbpagestyle{empty} 

\usepackage{amsthm}
\usepackage{algorithmic}
\usepackage[ruled,vlined,linesnumbered]{algorithm2e}
\usepackage{booktabs}
\usepackage{multirow}
\usepackage{supertabular}
\usepackage{enumitem}
\usepackage{graphicx, subfig}
\usepackage{balance}

\newtheorem{theorem}{Theorem}[section]

\newtheorem{problem}{Problem}
\newtheorem{property}[theorem]{Property}

\theoremstyle{definition}
\newtheorem{definition}{Definition}[section]
\newtheorem{example}{Example}[section]
\newtheorem{remark}{Remark}

\begin{document}
\title{Reliable Community Search in Dynamic Networks}

\author{Yifu Tang}
\affiliation{%
  \institution{Deakin University}
  \city{Geelong}
  \state{Australia}
}
\email{tangyif@deakin.edu.au}

\author{Jianxin Li}\authornote{Corresponding author}
\affiliation{%
  \institution{Deakin University}
  \city{Geelong}
  \country{Australia}
}
\email{jianxin.li@deakin.edu.au}

\author{Nur Al Hasan Haldar}
\orcid{0000-0001-5109-3700}
\affiliation{%
  \institution{The University of Western Australia}
  \city{Perth}
  \country{Australia}
}
\email{nur.haldar@uwa.edu.au}

\author{Ziyu Guan}
\affiliation{%
  \institution{Xidian University}
  \city{Xi'an}
  \country{China}
}
\email{zyguan@xidian.edu.cn}

\author{Jiajie Xu}
\affiliation{%
  \institution{Soochow University}
  \city{Suzhou}
  \country{China}
}
\email{xujj@suda.edu.cn}

\author{Chengfei Liu}
\affiliation{%
  \institution{Swinburne University of Technology}
  \city{Melbourne}
  \country{Australia}
}
\email{cliu@swin.edu.au}

\begin{abstract}
Searching for local communities is an important research problem that supports advanced data analysis in various complex networks, such as social networks, collaboration networks, cellular networks, etc. 
The evolution of such networks over time has motivated several recent studies to identify local communities in dynamic networks. 
However, these studies only utilize the aggregation of disjoint structural information to measure the quality and ignore the reliability of the communities in a continuous time interval.
To fill this research gap, we propose a novel $(\theta,k)$-$core$ reliable community (CRC) model in the weighted dynamic networks, and define the problem of \textit{most reliable community search} that couples the desirable properties of connection strength, cohesive structure continuity, and the maximal member engagement.
To solve this problem, we first develop a novel edge filtering based online CRC search algorithm that can effectively filter out the trivial edge information from the networks while searching for a \textit{reliable} community. 
Further, we propose an index structure, Weighted Core Forest-Index (WCF-index), and devise an index-based dynamic programming CRC search algorithm, that can prune a large number of insignificant intermediate results and support efficient query processing.
Finally, we conduct extensive experiments systematically to demonstrate the efficiency and effectiveness of our proposed algorithms on eight real datasets under various experimental settings.
\end{abstract}

\maketitle

\pagestyle{\vldbpagestyle}
\begingroup\small\noindent\raggedright\textbf{PVLDB Reference Format:}\\
\vldbauthors. \vldbtitle. PVLDB, \vldbvolume(\vldbissue): \vldbpages, \vldbyear.\\
\href{https://doi.org/\vldbdoi}{doi:\vldbdoi}
\endgroup
\begingroup
\renewcommand\thefootnote{}\footnote{\noindent
This work is licensed under the Creative Commons BY-NC-ND 4.0 International License. Visit \url{https://creativecommons.org/licenses/by-nc-nd/4.0/} to view a copy of this license. For any use beyond those covered by this license, obtain permission by emailing \href{mailto:info@vldb.org}{info@vldb.org}. Copyright is held by the owner/author(s). Publication rights licensed to the VLDB Endowment. \\
\raggedright Proceedings of the VLDB Endowment, Vol. \vldbvolume, No. \vldbissue\ %
ISSN 2150-8097. \\
\href{https://doi.org/\vldbdoi}{doi:\vldbdoi} \\
}\addtocounter{footnote}{-1}\endgroup

\ifdefempty{\vldbavailabilityurl}{}{
\vspace{.3cm}
\begingroup\small\noindent\raggedright\textbf{PVLDB Artifact Availability:}\\
The source code, data, and/or other artifacts have been made available at \url{\vldbavailabilityurl}.
\endgroup
}

\input{Introduction}
\input{Problem_Definition}
\input{Online_Search}
\input{Index_Search}
\input{Index_Const}
\input{Experiment}

\input{Related_Works}
\input{Conclusion}

\begin{acks}
 This work was mainly supported by the ARC Linkage Project under Grant No. LP180100750.
\end{acks}


\balance
\bibliographystyle{ACM-Reference-Format}
\bibliography{final}

\end{document}

%% file: Introduction.tex

\section{Introduction}
\label{Sec:Intro}
Local community search has attracted much attention in recent years and has shown its great success in different applications, e.g., personalized recommendation \cite{interdonato2017personalized, li2020personalized}, destination marketing \cite{park2008collaboration}.
In general, local community search aims to identify a densely connected structure with regard to a query vertex.
Majority of the existing works \cite{clauset2005finding,luo2008exploring,luo2020local} on local community search consider static network structure. 
For instance, Clauset et al.~\cite{clauset2005finding} and Luo et al.~\cite{luo2008exploring} proposed local modularity to measure the quality of the community in the static network. 
A random walk based community detection model from multiple static networks is proposed in ~\cite{luo2020local}. 
Some existing works~\cite{bu2015local,ditursi2017local} take into consideration that network structure may change over time and propose local community search in dynamic networks. 
For example, Bu et al.~\cite{bu2015local} provides a modularity-based criterion to find local communities in a dynamic network and update them in an incremental manner by monitoring the changes. 
In another work, DiTursi et al.~\cite{ditursi2017local} discovered the dynamic communities with optimal time intervals by minimizing \textit{temporal conductance}, a well-known metric to measure the quality of a community.
However, the above-mentioned works assess the community quality using their aggregated structural cohesiveness at independent timestamps and ignore the evolving structure of a community over time. 
In the dynamic network, the continuity of the community cohesiveness is an important factor in determining whether a community is reliable. 
For example, it is desirable to hire a team that continuously delivers high-quality outputs together over time in the collaboration network. 
Finding the user groups with a longer duration of reacting to the social event can help better analyze user behavior on social media.
Existing works largely ignore the continuity of the community cohesiveness.
In addition, they did not consider the edge weight, e.g., connection strength between a node pair, which incurs the new computational challenge to solve local community search in dynamic networks. 

To fill this research gap, we propose a novel community model of $(\theta,k)$-$core$ reliable community in dynamic networks where: (i) the community is a $k$-$core$ with each edge weight no less than the weight threshold $\theta$, and (ii) spans over a period of time. The most reliable local community search aims to find the community with the maximum reliability score, which is defined by coupling temporal continuity and member engagement. 
In other words, this work jointly models the three important properties, i.e., connection strength, cohesiveness continuity, and member engagement, of a community in a dynamic network.

\begin{figure}
    \centering
    \includegraphics[width=0.4\textwidth]{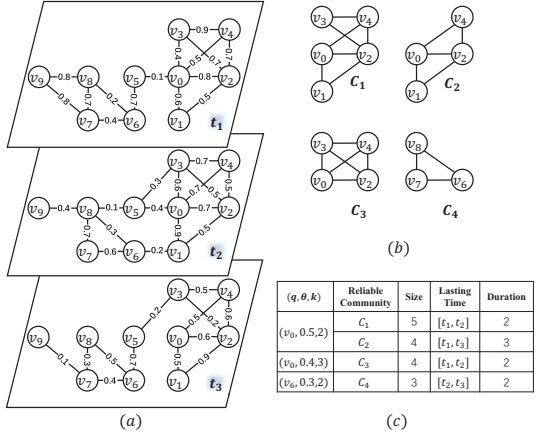}
    \caption{Interactions in a small graph of three timestamps.}
    \label{fig:intro}
\end{figure}

\begin{example}

Figure~\ref{fig:intro} shows an illustrative example of a dynamic online social network of three timestamps $t_1, t_2, t_3$, where the weight of an edge represents the connection strength.
Assume that a cosmetics company wants to post ads for a new product that can attract customers from local communities. 
The effective way is to search for a community from an existing customer, where the members might have continuous interaction or communication with each other with a minimum connection strength.
Our proposed $(\theta, k)$-$core$ reliable community search model can support such an application. 
Suppose, $v_0$ is the target user and $k$=3, $\theta$=0.4 are the query inputs. We can identify the community $C_3$=$\{v_0,v_2,v_3,v_4\}$ as the optimal result, which has four members and spans for two continuous timestamps $t_1$ and $t_2$. 
Similarly, it returns two communities $C_1$ and $C_2$ when $k$=2 and $\theta$=0.5.
Among these two communities, $C_2$ can be more acceptable than $C_1$ for the cosmetic company. This is because the members of $C_2$ kept their close interaction for 3 continuous timestamps and the size of $C_2$ is considerably similar to $C_1$. This enables the company to identify ads placeholder easily and cost-effectively that can maximize the impact of their ads. 
\end{example}

Although there exist some similar works \cite{li2018persistent, li2021efficient, 9001192, lin2021mining} that identify meaningful communities over time, our proposed problem is more acceptable due to more sophisticated research challenges. 
In \cite{li2018persistent}, Li et al. defined the persistent community search as the maximal $k$-$core$ where each vertex's accumulated degree meets the $k$-$core$ requirement within a time interval. 
They designed a novel temporal graph reduction algorithm and searched the maximum persistent community utilizing pruning and bounding techniques, which takes exponential complexity. 
Expanding from \cite{li2018persistent}, Li et al. \cite{li2021efficient} studied a single query vertex based persistent community search 
by developing an enumeration-based subgraph search algorithm. 
In \cite{9001192}, Qin et al. proposed the stable communities by first selecting the centroid vertices where each centroid vertex has a certain number of neighbors with the desired similarity, and the star-shape of the centroid vertex and its neighbors exists frequently in a period of time; and then clustering the network vertices into stable groups based on the selected centroids. 
Similar to \cite{9001192}, Lin et al. \cite{lin2021mining} defined frequency-based dense subgraphs that satisfy the quasi-clique structure with at least $\theta$ vertices and the degree of each vertex exceeds a given threshold. The proposed searching algorithm takes exponential complexity.
However, the persistent community in \cite{li2018persistent, li2021efficient} did not consider the weight of the edge and its time complexity is too high for dealing with large-scale networks. The frequency-based subgraph in \cite{9001192, lin2021mining} ignored the continuity of the cohesive structure and failed to maximize member engagement.

To solve the proposed problem of the most \textit{reliable} local community search, a naive idea is to enumerate all the possible community candidates and select the satisfied results by checking their edge weight and duration. 
However, this may incur an exponential time cost. 
To address the computational challenge, in this paper, we firstly propose an efficient eligible edge filtering online search algorithm that utilizes the minimum edge requirement of $k$-$core$ to compute the reliability upper bound to prune a large number of edge sets without probing their corresponding ($\theta$,$k$)-$core$ community candidates.
To further accelerate the query processing, we develop a weighted core forest index by maintaining the standard $\theta$-threshold values and the $(\theta,k)$-$core$ structural information of vertices, which supports efficient retrieval of $k$-$core$ with regard to different thresholds and timestamps. 
Following this, we design an index-based dynamic programming algorithm, and derive the reliability upper bound of communities w.r.t. the time interval during the dynamic programming procedure to avoid probing the unsatisfactory community candidates. 
Besides, the index construction, maintenance, and compression are well presented in this paper.

The main contributions of this work are as below:
\begin{itemize}
    \item We propose a novel problem of the most reliable community search that jointly considers community continuity, community size, and connection strength for online network analysis services.
    \item We develop an efficient online search algorithm by deriving and applying the properties of pruning the ineligible edges w.r.t. the given query conditions. 
    \item We further present a weighted core forest index and develop an index-based dynamic programming algorithm to solve the most reliable community search problem in a more efficient way. 
    \item We conduct extensive experiments to show the efficiency and effectiveness of the proposed algorithms and community model by using eight real-world datasets and comparing with three existing studies.
\end{itemize}

The remainder of this paper is organized as follows. 
First, we formalize the most reliable local community search problem in Section~\ref{Sec:ProbDef} and develop the online search algorithm in Section~\ref{Sec:Online}.
Then, we introduce our index structure and the detailed index-based search algorithm in Section~\ref{Sec:Index}.
The procedures of index construction, maintenance and compression are shown in Section~\ref{Sec:Index-plus}.
Experimental evaluation and results are discussed in Section~\ref{sec:exp}.
Finally, we discuss the related work in Section~\ref{Sec:RelatedWork} and conclude the work in Section~\ref{Sec:Conclusion}.

%% file: Problem_Definition.tex
\section{Preliminaries and Problem Definition}
\label{Sec:ProbDef}

In this section, we first present the preliminaries and then formalize the problem of the most reliable local community search. 
\begin{definition}[Dynamic Networks]
	A dynamic network $\mathcal{G} = \{G_{t_1}, ..., G_T\}$ is a sequence of time-variant weighted graph instances $\{G_{t_1}, ..., G_T\}$ s.t., $t_1<t_2<...<T$, where each timestamped instance $G_{t} = (V_t, E_t, W_t)$ contains a set of vertices $V_t$, a set of edges $E_t$ with the weights $W_t(e)\in (0,1]$ for $\forall e \in E_t$.  
\end{definition}
In this work, we ignore the isolated vertices, so that vertex updates can be supported by edge insertions and deletions. The edge insertion with a new endpoint can represent the vertex addition and edge deletion isolated an endpoint reflects the vertex deletion.
For simplicity, we assume all graph instances share a fixed set of vertex $V$, i.e. $G_{t} = (V, E_t, W_t)$. 
The edge weight is a widely-used network feature to represent the interaction frequency, similarity, or connection strength between vertices.
In this work, we normalize the edge weight to be in $(0,1]$.

For a graph instance $G_t=(V, E_t,W_t)$, $deg(u, G_t)$ denotes the degree of a vertex $u$ in $G_t$, which is the number of neighbors of $u$ in $G_t$.
Like \cite{seidman1983network}, we consider $k$-$core$ in $G_t$ as a connected subgraph $G_t^k=(V^k, E_t^k)$ where each vertex has the degree no less than $k$, i.e. $\forall u\in V_t^k, deg(u, G_t^k)\geq k$.

\begin{definition}[($\theta$,$k$)-$core$]
Given a graph instance $G_t=(V, E_t, W_t)$, an integer $k$, and a threshold $\theta$, a connected subgraph $G_t^{\theta,k}=(V^{\theta,k}, E_t^{\theta,k}, W_t^{\theta,k})$ is called a ($\theta$,$k$)-$core$ of $G_t$ if $G_t^{\theta,k}$ is a $k$-$core$ and each edge has the weight no less than $\theta$, i.e. $\forall u\in V^{\theta,k}, deg(v, G^{\theta,k})\geq k$ and $\forall e\in E_t^{\theta,k}, W_t^{\theta,k}(e)\geq \theta$.
\end{definition}

\begin{definition}[Time Interval based ($\theta,k$)-$\underline{c}ore$ \underline{R}eliable \underline{C}ommunity (CRC)]\label{def:reliablecommunity}
Given a dynamic network $\mathcal{G}$ = $\{G_{t_1}, ..., G_T\}$, an integer $k$, a threshold $\theta$, and a time interval $T_C=[t_s,t_e]$, 
a ($\theta,k$)-$core$ reliable community is a subgraph $C=(V_C, E_C)$ that spans continuously from $t_s$ to $t_e$ and for each timestamp $t_n\in T_C$, the subgraph induced by $E_C$ from the graph instance $G_{t_n}$ is a ($\theta$,$k$)-$core$, i.e. $\forall t_n\in T_C$, $G_{t_n}[E_C]$ is a ($\theta$,$k$)-$core$ of $G_{t_n}$. In the remainder of this work, ($\theta,k$)-$core$ reliable community is called CRC for brevity.
\end{definition}

Based on Definition~\ref{def:reliablecommunity}, a CRC maintains a cohesive structure with the required connection strength of a time interval in the dynamic network. Its reliability score can be measured by coupling the continuity and size of the community as below:

\begin{definition}(Reliability Score of CRC)
    \label{def:reliabilityScore}
    Given a dynamic network $\mathcal{G}=\{G_{t_1}, G_{t_2}, ...,G_T\}$, a query time interval $T_Q=[t_i,t_j]$, and a CRC $C=(V_C, E_C)$ with regard to the time interval $T_C=[t_s,t_e]$, where $T_C\subseteq T_Q$, the reliability score $S_{rel}(C)$ is defined as the harmonic average of its normalized duration and size:
    \begin{align}
        \label{eq:rel}
            S_{rel}(C)=(1+\alpha^2)\cdot\frac{\mathcal{N}(V)\cdot\mathcal{N}(T)}{(\alpha^2\cdot\mathcal{N}(V))+\mathcal{N}(T)}
        \end{align}
    where $\mathcal{N}(V)=|V_C|/|V^k_{max}|$ and $\mathcal{N}(T)=|T_C|/|T_Q|$ represent the normalized size and duration length respectively. 
    $V^k_{max}$ denotes the $k$-$core$ with the maximum size in $G_t$ ($t\in T_Q$).
\end{definition}

From Equation~\ref{eq:rel}, we can infer that reliability score is monotonically related to size and duration of the community. Parameter $\alpha$ means that community duration is $\alpha$ times as much important as community size, so that higher $\alpha$ tends to find the community with longer duration and smaller size. The default $\alpha=1$ means both community size and duration are important equivalently.

Based on the above definitions and reliability measurement, we formalize the most reliable local community search problem as below.
\begin{problem}[Most Reliable Local Community Search]\label{problem_def}
	Given a dynamic network $\mathcal{G}=\{G_{t_1}, G_{t_2}, ...,G_T\}$, a query vertex $q$, a threshold $\theta$, a structural constraint integer $k$, and a query time interval $T_Q=[t_i,t_j]$, the problem of the Most Reliable Local Community Search is to find the CRC $C=(V_C,E_C)$ and its continuous time interval $T_C=[t_s,t_e]$, satisfying 
	\begin{align}
    \label{eq:opt}
        \arg\max_{C\subseteq V}S_{rel}(C)
    \end{align}
    subject to 
        $q\in V_C$; $\forall v\in V_C$, $deg(v, C)\geq k$; $\forall e\in E_C \land \forall t\in T_C$, $W_t(e)\geq \theta$; and $T_C\subseteq T_Q$.
    \end{problem}
As shown in Figure~\ref{fig:intro}, when the query time interval is $[t_1,t_3]$, we can obtain two reliable communities $C_1$ and $C_2$ w.r.t. the query input $(v_0,0.5,2)$. 
The maximal $2$-$core$ is composed of 10 vertices. When $\alpha=1$, we have $S_{rel}(C_1) =S_{rel}(C_2)=0.57$. When $\alpha$ increases to $2$, we have $S_{rel}(C_1)=0.63<S_{rel}(C_2)=0.77$, and $C_2$ with longer duration becomes the optimal result.

To solve the most reliable community search problem, a naive solution is to compute all the $(\theta,k)$-$core$ at each timestamp, and then verify their longest duration in the dynamic network.
After that, their reliability scores can be obtained by multiplying their size and the number of continuous timestamps. Finally, the most reliable community can be returned by selecting the ones with the maximum reliability scores. 
However, the operation of finding all the $(\theta,k)$-$core$ needs to probe all the combinations of edges that form a connected subgraph with no less than $k(k+1)/2$ edges and $k+1$ vertices, i.e., satisfying the conditions of minimal $k$-$core$ component. Thus, we can remark that the computational cost of finding the most reliable community is in exponential complexity. 

%% file: Online_Search.tex
\section{Online Reliable Community Search}
\label{Sec:Online}

To efficiently solve the problem of reliable local community search, in this section, we will present a novel \textit{\underline{E}ligible \underline{E}dge \underline{F}iltering} (\textit{EEF}) based Online CRC Search algorithm. 
Different from the naive idea discussed in Section~\ref{Sec:ProbDef}, \textit{EEF} does not need to generate all the $(\theta,k)$-$core$ candidates, which can greatly reduce the query time cost.

Given a graph instance $G_{t_n}=(V,E_{t_n})$ and a weight threshold $\theta$, we can filter out ``ineligible'' edges whose weights are less than $\theta$ and maintain only the ``eligible'' edges.
We identify the set of eligible edges, denoted by $E_{t_n,\theta}$, i.e. $E_{t_n,\theta}=\{e\in E_{t_n}|W_{t_n}(e)\geq\theta\}$ for the CRC construction.

\begin{definition}[Eligible Lasting Time of Edge]
\label{def:lasttime}
Given a graph instance $G_{t_n} = (V, E_{t_n})$ at timestamp $t_n\in[t_i,t_j]$, and a threshold $\theta$, for an edge $e\in E_{t_n\theta}$, its eligible lasting time $\lambda_{t_n,\theta}(e)$ is measured by the length of the longest time interval $[t_m,t_n]$ $(t_i\leq t_m\leq t_n])$ if $W_{t_x}(e) \geq \theta$ for $\forall t_x \in$ [$t_m$, $t_n$]. 
\end{definition}

Eligible lasting time calculates the number of continuous timestamps that the edge is ``eligible'' until the current timestamp.
For example, in Figure~\ref{fig:intro}, for edge $e=(v_0,v_1)$, we have $\lambda_{t_1,0.6}(e)=1$, $\lambda_{t_2,0.6}(e)=2$ and $\lambda_{t_3,0.6}(e)=0$.
We can easily derive that the eligible lasting time of edges can be incrementally computed by accessing the dynamic network chronologically.

Based on the eligible time, we can easily identify common edges of multiple continuous graph instances, that can be utilized to construct CRC with varying duration.
Different from the vertex-induced subgraph, the subgraph induced by an eligible edge set provides the guarantee to meet the requirement of edge weight $\theta$. Therefore, the eligible edge set can be used to prune the unqualified $k$-$core$ candidates by using the following property.

\begin{property}[Minimum $k$-$core$]
	\label{prop:minkCore}
Given an edge set $E_{t_n,\theta}$ at timestamp $t_n$ with regards to a threshold $\theta$, $E_{t_n,\theta}$ can be pruned without probing its induced communities if $|E_{t_n,\theta}| < k(k+1)/2$, i.e., the number of edges does not meet the density requirement of $k$-$core$. 
\end{property}

In addition, we can calculate the upper bound of the reliability score of CRCs constructed using $E_{t_n,\theta}$. 
Given $G'=(V',E')$ as a $k$-$core$ in the induced subgraph $G_{t_n}[E_{t_n,\theta}]$, $|E_{t_n,\theta}|\geq|E'|\geq(k\cdot|V'|)/2$ must hold because there are at least $k$ edges for a vertex in the $k$-$core$. 
Therefore, if $G'$ can form a CRC with its duration as $d$, then its vertex size satisfies that $|V'|\leq2|E_{t_n,\theta}|/k$.
Thus, we have  $\mathcal{N}(V')\leq\frac{2|E_{t_n,\theta}|/k}{|V^k_{max}|}$ and $\mathcal{N}(T)=d/|T_Q|$ and we can determine the upper bound of reliability score of the CRC constructed by the given eligible edge set.

For simplicity of presentation, we omit the normalizers $|V^k_{max}|$ of the maximum community size and $|T_Q|$ of the query time interval in the following equations.

\begin{property}[$S_{rel}$ Upper Bound of CRC w.r.t. $E_{t_n,\theta}$]
\label{prop:ub1}
Given an eligible edge set $E_{t_n,\theta}$, an integer $k$, the upper bound reliability score $UBR_{t_n}^d$ of the CRC constructed using $E_{t_n,\theta}$ whose duration is $d$ is calculated as:
\begin{align}
\label{eq:ub}
    UBR^d_{t_n} = (1+\alpha^2)\cdot\frac{2|E_{t_n,\theta}|/k\cdot d}{(\alpha^2\cdot2|E_{t_n,\theta}|/k)+d}
\end{align}
\end{property}

The key idea of \textit{EEF}-based Online CRC Search is to filter out edges in each graph instance $G_{t_n}$ using the given threshold $\theta$ while maintaining the lasting time of each edge by a timestamp.
As shown in Algorithm~\ref{algo:edge}, we first initialize $C_{opt}$ and $maxS$ to store the most reliable community and its reliability score (line 1). 
Then, for each timestamp $t_n$, we traverse the edges of $G_{t_n}$ starting from the query vertex $q$ in \textit{Breadth-First Search} manner.
In the meantime, vertices and edges that violating the degree and weigh constraints are pruned.
During the traversing, the eligible time of edges is updated incrementally, and the eligible edges are added to the edge set $E_{t_n,\theta}$.
Having $E_{t_n,\theta}$, we calculate its upper bound $UBR^1_{t_n}$ of potential CRC whose duration is 1 (lines 2-10).
Then, we visit each timestamp $t_n$ in the descending order of $UBR^1_{t_n}$, which provides a best-first search strategy to exploit the CRCs.
We utilize $E_{t_n,\theta}$ to construct CRC with duration $d$ iterating from 1 to $|[t_i,t_n]|$ (lines 11-19).
At each iteration, we select the edge set $E'$ where each edge has the eligible time no less than $d$ and update the upper bound w.r.t. $d$.
Then we adopt Property~\ref{prop:minkCore} and Property~\ref{prop:ub1} to prune the CRC construction if $|E'|$ is too small or $UBR^d_{t_n}$ cannot exceed $maxS$.
After that, we can extract the CRC $C$ (i.e., local maximal $k$-$core$) from the induced subgraph $G_{t_n}[E']$ by finding the connected component containing $q$ after the core decomposition, and then update $C_{opt}$ and $maxS$.
Finally, the algorithm returns $C_{opt}$ as the optimal result.
\begin{algorithm}[ht]
  \caption{\textit{EEF}-based Online CRC Search}\label{algo:edge}
  \KwIn{A dynamic weighted graph $\mathcal{G}=\{G_{t_1}, G_{t_2},...\}$, an integer $k$, a query vertex $q$, a query time period $T_Q=[t_i,t_j]$, and a threshold $\theta$.}
  \KwOut{The Most Reliable Community $C_{opt}$}
  $C_{opt}\gets \emptyset$; $maxS\gets0$\;
  \For{$t_n\gets t_i$ to $t_j$}{
  \While{traverse edges $e=(v,w)$ starting from $q$ in \textit{BFS}}{
    \lIf{$deg(v,G_{t_n})\leq k$}{Remove vertex $v$ from $G_{t_n}$} 
    \If{$W_{t_n}(e)\geq\theta$}{
        $E_{t_n,\theta}\gets E_{t_n,\theta}\cup \{e\}$\;
        \If{$e\in E_{t_{n-1},\theta}$}{
            $\lambda_{t_n,\theta}(e)\gets \lambda_{t_{n-1},\theta}(e)+1$
            }
    }
    \lElse{Remove edge $e$ from $G_{t_n}$}
 }
    Calculate $UBR^1_{t_n}$ by Equation~\ref{eq:ub}
}
\For{$t_n\in T_Q$ in descending order of $UBR^1_{t_n}$}{
    \For{$d\gets 1$ to $|[t_i,t_n]|$}{ \label{a1:3for}
        $E'\gets \{e\in E_{t_n,\theta}|\lambda_{t_n,\theta}(e)\geq d\}$\;
        Calculate $UBR^d_{t_n}$ by Equation~\ref{eq:ub}\;
        \If{$|E'|\geq k(k+1)/2$ \& $UBR^d_{t_n}>maxS$}{
            $C\gets$ local maximal $k$-$core$ in $G_{t_n}[E']$\;\label{a1:3forend}
            \If{$S_{rel}(C)\geq S_{rel}(C_{opt})$}{
            $maxS\gets S_{rel}(C)$\;
            $C_{opt}\gets C$\; 
            }
        }
    }
 }
\textbf{Return} $C_{opt}$.
\end{algorithm}

The time complexity of Algorithm~\ref{algo:edge} can be analyzed as below. 
For each graph instance $G_{t_n}=(V,E_{t_n})$, it takes $O(|V|+|E_{t_n}|)$ to run the \textit{Breadth-First Search} that requires to visit very vertex and edge once. At the same time, the eligible time of each edge is obtained (lines 2-18).
Then, it needs $O(|[t_i,t_n]|\cdot|E_{t_n}|)$ to compute $|[t_i,t_n]|$ number of CRCs where each CRC is obtained by a core decomposition process that needs to consume $O(|E_{t_n}|)$ \cite{batagelj2003m} (lines 20-27).
Therefore, for the query interval of $|T_Q|$ timestamps, Algorithm~\ref{algo:edge} takes $O(\sum_{t_n\in T_Q}((|V|+|E_{t_n}|)+(|[t_i,t_n]|\cdot|E_{t_n}|)))$ in total, which can be rewritten as $O(|T_Q|\cdot((|V|+|E_t|)+(|T_Q|\cdot|E_t|))$, i.e., $O(|T_Q|^2\cdot|E_t| + |T_Q|\cdot(|E_t|+|V|))$, where $|E_t|$ denotes the average number of edges of the graph instances.

%% file: Index_Search.tex
\section{Index Based Reliable Community Search}
\label{Sec:Index}
To further accelerate the query processing, in this section, we first propose a forest index structure, called \textit{\underline{W}eighted \underline{C}ore \underline{F}orest Index} (\textit{WCF-Index}), to maintain the $(\theta,k)$-$core$ vertices for each graph instance $G_{t_n}$.
Then, we develop an index-based dynamic programming algorithm by using the proposed index and derive the reliable score upper bound with great pruning power to accelerate the query algorithm.
 
\subsection{WCF-Index}
\label{subsec:index}
The general idea of this index is to maintain the vertex candidates of the $(\theta,k)$-$core$ with regards to the given $\theta$ and $k$ at each timestamp, from which we can work out the satisfied CRCs containing $q$ with the different continuous time intervals.

\begin{definition}[$\theta$-threshold of a Vertex]
Given a graph instance $G_t$=$(V,E_t,W_t)$ at a timestamp $t$ and an integer $k$, for a vertex $u\in V$, it may have a set of $\theta$ values and their corresponding $(\theta, k)$-$core$ subgraphs containing $u$. Thus, we take the largest $\theta$ value in the $\theta$ set as the $\theta$-threshold of $u$, denoted as $\theta$-$thres_k(u,G_t)$.
\end{definition}

\begin{figure}[tb]
\begin{minipage}[t]{0.49\linewidth}
    \includegraphics[width=\linewidth]{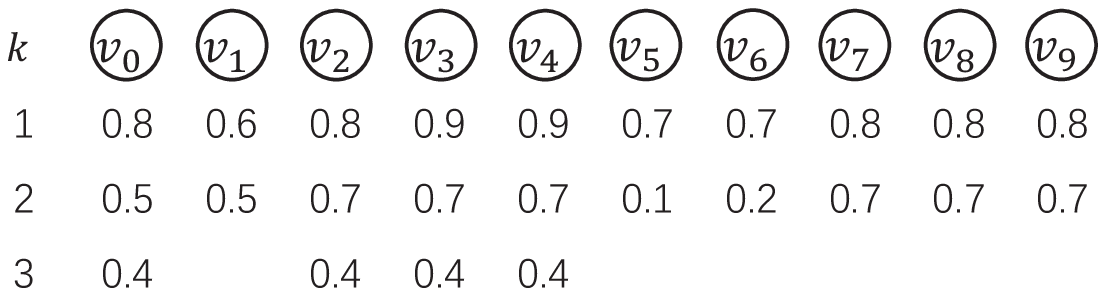}
    \caption{$\theta$-threshold of $G_{t_1}$} 
    \label{fig:theta-order}
\end{minipage}
\begin{minipage}[t]{0.49\linewidth}
    \includegraphics[width=\linewidth]{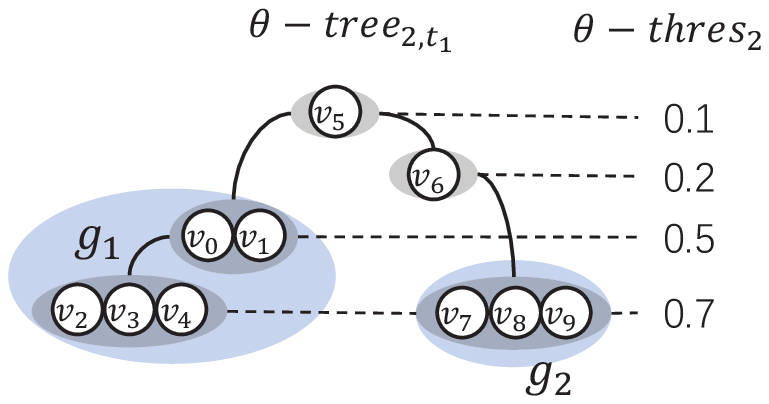}
    \caption{$\theta$-tree of $G_{t_1}$, $k$=2}
  \label{fig:theta-tree}
\end{minipage} 
\end{figure}

\begin{example}
Figure~\ref{fig:theta-order} shows the $\theta$-threshold of vertices in $G_{t_1}$ in Figure~\ref{fig:intro} (a) with regards to different $k$ values, e.g., $\theta$-$thres_2(v_1,G_t)$ is 0.5 because $(0.5,2)$-$core$  (i.e., $\{v_0, v_1, v_2, v_3, v_4\}$) exists in $G_{t_1}$, but no one $(\theta',2)$-$core$ containing $v_1$ exists if $\theta'>0.5$.
\end{example}

According to the above definition and the example, we are able to justify whether a vertex $v$ is contained in a $(\theta,k)$-$core$ for given $\theta$ and $k$ if the $\theta$-threshold values of vertices are maintained.
However, the $\theta$-threshold only implies the vertex candidatures of a $(\theta,k)$-$core$, but fails to reflect the structural connectivity of vertices. Therefore, it is highly desirable to design an index structure for maintaining the $\theta$-threshold and the structure information of vertices together.

Yang et al. in \cite{yang2019index} proposed a forest-based index to query $(k,\eta)$-$core$ in a static uncertain graph where $k$ implies the degree constraint of the vertex and $\eta$ implies the probability of the vertex to appears in the subgraph.
By maintaining $\eta$-$tree_k$ for each $k$, it can accelerate the search of all the $(k,\eta)$-$core$ with custom $\eta$ requirements.
Motivated by $\eta$-$tree_k$, in this work, we extend the concept of $\eta$-$tree_k$ to the dynamic weighted network to construct the $\theta$-$tree_{k,t}$ for each $k$ at time $t$, which can support quick retrieval of local maximal $(\theta,k)$-$core$ from the indexed graph instance $G_t$.

\begin{definition}[$\theta$-$tree_{k,t}$]
  Given a graph instance $G_t$, an integer $k$, $\theta$-$tree_{k,t}$ index is a tree structure, satisfying
   \begin{enumerate}
  \item \textbf{Node}: each tree node $\mathbb{V}$ is a set of maximal connected vertices in $G_t$ with same  $\theta$-threshold value, denoted as $\mathbb{V}.\theta$, i.e., $\forall v\in\mathbb{V}$, $\theta$-$thres_k(v,G_t)$ = $\mathbb{V}.\theta$;
  \item \textbf{Parent-child relationship}: for a node $\mathbb{W}$, $N_{\mathbb{W}}(G_t)$ denotes the tree nodes that are connected to $\mathbb{W}$ in $G_t$ with $\theta$-threshold smaller than $\mathbb{W}.\theta$. The parent node $\mathbb{V}$ of $\mathbb{W}$ is the node with the largest $\theta$-threshold in $N_{\mathbb{W}}(G_t)$, i.e. $\mathbb{V}=argmax_{\mathbb{V}\in N_{\mathbb{W}}(G_t)}\mathbb{V}.\theta$.
  \end{enumerate}
\end{definition}

\begin{example}
Figure~\ref{fig:theta-tree} presents the constructed $\theta$-$tree_{2,t_1}$ of $G_{t_1}$ from $\theta$-threshold of $G_{t_1}$ in Figure~\ref{fig:theta-order}.
If we search $(0.5,2)$-$core$ on $\theta$-$tree_{2,t_1}$, three tree nodes will be returned, i.e., $\{v_0, v_1\}$, $\{v_2, v_3, v_4\}$, and $\{v_7, v_8, v_9\}$. These tree nodes can induce two $(0.5,2)$-$core$, i.e., $g_1$ and $g_2$.
\end{example}

$\theta$-$tree$ can be composed of several trees where each tree represents a connected component in the graph instance. We denote $\mathcal{I}$ as the \textit{WCF-Index} where $\mathcal{I}[k][t]$ represents the $\theta$-$tree_{k,t}$ of each $k$ and $t$ in the dynamic network.

\begin{remark}
  \label{rmk1}
  In this work, we set the $\theta$-threshold as the standard values $\{0, 0.1, 0.2,...,0.9, 1\}$. If the $\theta$-threshold of a vertex is not in the standard set, we will round it down to the nearest standard value. 
  Accordingly, fetching $(\theta,k)$-$core$ with non-standard $\theta$ value will also be processed as the nearest rounded down standard value.
  For instance, to fetch $(\theta,k)$-$core$ with $\theta=0.55$, the index accesses the tree nodes from $\theta$-threshold of 0.5 and then examines the $\theta$-threshold of vertices in the tree node $\mathbb{V}$ if $\mathbb{V}.\theta<0.55$.
  In the following discussion, we skip this process for simplicity.
\end{remark}

\subsection{Dynamic Programming based CRC search}
\label{subsec:dyna}
To solve the most reliable community search problem, we need to compare CRC with different duration. 
In this section, we develop a dynamic programming algorithm based on the recursive relation of CRCs ending in consecutive timestamps and utilize the \textit{WCF-Index} to search CRC with varying duration efficiently.

Assume that the lasting time interval of the CRC is fixed (so does the duration), then we only need to extract the CRC with the largest size. 
Given the duration of the CRC is $d$ and the last timestamp it spans is $t_n$, we denote the maximal CRC w.r.t. the query input as $C(d,t_n)$.
We can easily derive the following recursive relation between CRCs:
\begin{align}
\label{eq:inter}
    C(d,t_n)\subseteq C(d-1,t_{n-1})\cap C(d-1,t_n)
\end{align}
The base situation is $C(1,\cdot)$ that can be retrieved from \textit{WCF-Index}.
Based on Eq.~\ref{eq:inter}, we can devise a DP algorithm to compute $C(d,t_n)$.
More specifically, to get $C(d,t_n)$, we simply compute the intersection of $C(d-1,t_{n-1})$ and $C(d-1,t_n)$ and extract the local maximal $(\theta,k)$-core using core decomposition.
The intermediate result of $C(d,t_n)$ with varying $d$ is maintained to support the adoption of the dynamic programming.

If at a timestamp $t_a$, the maximal community $C(1,t_a)$ does not exist, i.e. for a given query, there is no such subgraph satisfying $(\theta,k)$-$core$ constraint at time $t_a$, then it implies that further calculations depending on $C(1,t_a)$ are unnecessary. In this work, these kinds of timestamps like $t_a$ are called anchored timestamps of a query time interval. 
The anchored timestamps split the query interval $[t_i,t_j]$ into several non-overlapping time intervals $T_S=\{T_1, T_2,...\}$. 
For each interval $T_i\in T_S$, we can compute the upper bound of the reliability score of the communities.

\begin{property}[$S_{rel}$ Upper Bound of CRC w.r.t. $T_i$]
\label{prop:wcf_ub}
Given a time interval $T_i=[t_s,t_e]$ where $C(1,t_n)$ exists for every $t_n\in[t_s,t_e]$, we can construct an array $M=(\mu_s,\mu_{s+1},...\mu_e)$ to store the size of $C(1,\cdot)$, where $\mu_n$ denotes the size of $C(1,t_n)$.
The upper bound reliability score (UBR) in this time interval can be calculated by:
\begin{align}
\label{eq:ub2}
    UBR_{T_i} = max_{\mu_n\in M}((1+\alpha^2)\cdot\frac{\mu_n\cdot LCT(\mu_n,M)}{(\alpha^2\cdot\mu_n)+LCT(\mu_n,M)})
\end{align}
where $LCT(\mu_n,M)$ stands for the length of the \underline{L}ongest \underline{C}onsecutive \underline{T}imestamps $S$ of $\mu_n$ in array $M$ such that $\{i\in S|\mu_i\leq\mu\}$. 
\end{property}

\begin{figure}
  \centering
  \includegraphics[width=0.46\textwidth]{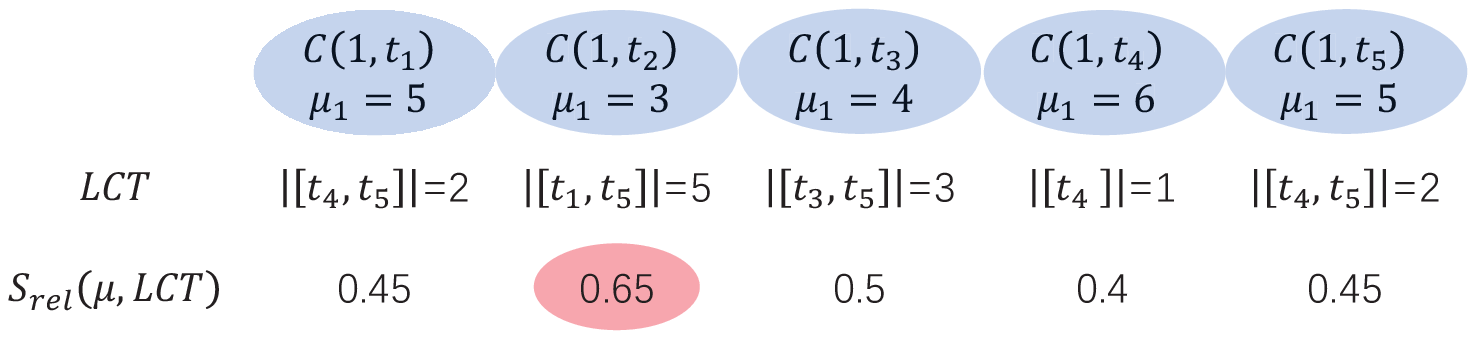}
  \caption{$UBR$ of $[t_1,t_5]$ }
  \label{fig:ubr}
\end{figure}

\begin{example}
\label{exp:4.3}
For a community $C(1,t_n)$ of size $\mu_n$, the largest reliability score of a CRC constructed by $C(1,t_n)$ is determined by the longest possible duration that $\mu_n$ can remain. Hence, the upper bound score of a time interval is the maximum value among all the largest possible scores for each $C(1,\cdot)$ size.
Figure~\ref{fig:ubr} shows an illustrative example of calculating the UBR of interval $[t_1,t_5]$ where we assume $|V^k_{max}|=10$. The size of $C(1,t_3)$ is 4 and the longest continuous timestamps for this size is 3 ($[t_3,t_5]$), so the largest possible score of the CRC constructed by $C(1,t_3)$ is $0.5/10*4+0.5/5*3=0.5$. The largest possible reliability score is obtained by a CRC constructed by $C(1,t_2)$ that contains three vertices and spans for five timestamps.
\end{example}

\begin{remark}
Similar to Property 2, we use $UBR^d_{T_i}$ to denote the upper bound calculated by the size of $C(d,\cdot)$.
However, it can determine the maximum reliability score of the community with duration longer than $d$. We can derive that
\begin{align}
\label{eq:ub3}
    UBR^d_{T_i} = max_{\mu_n\in M}(\frac{(1+\alpha^2)\cdot\mu_n\cdot (d+LCT(\mu_n,M)-1)}{(\alpha^2\cdot\mu_n)+(d+LCT(\mu_n,M)-1)})
\end{align}
where $\mu_n$ denotes the size of $C(d,t_n)$ and $(LCT(\mu_n,M)-1)$ represents the additional lasting timestamps of size $\mu_n$ on top of $d$.
In the process of community search, $UBR_{T_i}$ can be updated with different duration of CRC and provide sustainable pruning power.
Having $UBR_{T_i}$ calculated for each interval $T_i$ and updated during the community exploration, we can skip exploring \textit{CRCs} if $UBR_{T_i}$ is no larger than the reliability score of intermediate community candidates we have obtained during the query processing.
\end{remark}

Algorithm~\ref{algo:index_query} presents the detailed  dynamic programming procedure of the \textit{WCF-Index} based CRC Search. 
We first initialize a table $L_C$, $maxS$, and $C_{opt}$ to store the extracted communities, the maximum reliability score, and the most reliable community (line 1).
For each timestamp $t_n\in[t_i,t_j]$, we can obtain $C(1,t_n)$ from the $\theta$-$tree_{k,t_n}$ index, and store it in $L_C[1][n]$ (lines 2-6). In addition, we also determine whether $t_n$ is an anchored timestamp.
Then we split $T_Q$ into several non-overlapping time intervals $T_S=\{T_1, T_2,...\}$ with valid $C(1,\cdot)$ by the anchored timestamps, and calculate their upper bound reliability score (lines 7-8).
For each individual time interval $T_i=[t_s, t_e]$, if its upper bound is no larger than $maxS$, then the time interval is pruned (line 10). 
Otherwise, we initialize an array $Q$ to store the size of CRC and compute the CRC with various duration $d$ based on Eq.~\ref{eq:inter}, and store the intermediate CRC $C(d,t_x)$ in $L_C[d][x]$ (lines 11-16).
Then, we update $maxS$ and $C_{opt}$ and add the size of $C(d,t_x)$ to $Q$ for upper bound calculation (lines 17-20).
Once $C(d,\cdot)$ has been explored for every $t_x\in[t_s,t_e]$, we can update the $UBR$ and determine whether it is necessary to explore communities with longer duration in this interval (line 21-22).
Finally, the algorithm returns the most reliable community $C_{opt}$ whose reliability score is the largest.

\begin{algorithm}
  \KwIn{A dynamic weighted graph $\mathcal{G}=\{G_{t_1}, G_{t_2},...G_{T}\}$, query time $[t_i,t_j]$, the \textit{WCF-Index} $\mathcal{I}$, integer $k$, weight threshold $\theta$, query vertex $q$}
  \KwOut{the most reliable community $C_{opt}$}
  $L_C\gets [[]]$; $maxS\gets0$; $C_{opt}\gets\emptyset$\;
  \For{$t_n\in [t_i,t_j]$}{
    Extract $C(1,t_n)$ from $\mathcal{I}[k][t_n]$\;
    \If{$C(1,t_n)$ is $\emptyset$}{set $t_n$ as anchored timestamp\;}
    \lElse{$L_C[1][n]\gets C(1,t_n)$}
  }
  Get all the consecutive timestamps $T_S=\{T_1, T_2,...\}$ split by the anchored timestamp\;
  Calculate upper bound $ub=\{UBR^1_{T_1}, UBR^1_{T_2},...\}$ for each consecutive time sequence by Equation ~\ref{eq:ub2}\;
  \For{$T_i=[t_s,t_e]\in T_S$ by descending of $UBR_{T_{i}}$}{
    \lIf{$UBR^1_{T_{i}}\leq maxS$}{
        \textbf{continue}
    }
        \For{$d\gets 1$ \textbf{to} $|[t_s,t_e]|$}{
            $M\gets[]$\;
            \For{$t_x\in [t_s,t_e]$}{
                \If{$d\leq |[t_s,t_x]|$}{
                    \If{$d>1$}{
                      $L_C[d][x]\gets $local maximal $k$-core in $L_C[d-1][x-1]\cap L_C[d-1][t_x]$\;
                  }
                $maxS\gets max(S_{rel}(L_C[d][x]),maxS)$\;
                \If{$S_{rel}(L_C[d][x])\geq S_{rel}(C_{opt})$}{
                $C_{opt}\gets L_C[d][x]$\; 
                }
                $M.append(|L_C[d][x]|)$\;
                }
              }
            Calculate $UBR^d_{T_i}$ by Equation ~\ref{eq:ub3}\;
            \lIf{$UBR^d_{T_{i}}\leq maxS$}{
                \textbf{break}
            }
        }
  }
  \textbf{Return} $C_{opt}$
  \caption{\textit{WCF-Index} based CRC Search}
  \label{algo:index_query}
  \end{algorithm}

The time complexity of Algorithm~\ref{algo:index_query} is dominated by the operation of finding CRCs with various duration (lines 10-23) as the $C(1,\cdot)$ community can be queried from the index in constant time. 
In the worst case, there are up to $|T_Q|^2$ subgraphs to be explored and the community construction takes $O(|E_t|)$ complexity, where $T_Q$ is the query interval and $|E_t|$ is the average number of edges of graph instance $G_t$.
The total complexity is $O(|T_Q|^2\cdot|E_t|)$. Compared with the EEF-based Online CRC Search Algorithm, \textit{WCF-Index} can avoid searching a large number of edges, which helps to reduce the time cost of computing $C(1,\cdot)$.

%% file: Index_Const.tex
\section{WCF Index Construction, Maintenance, and Compression}
\label{Sec:Index-plus}
In this section, we describe the procedure of index construction, and propose index maintenance and compression strategies to support efficient query processing over dynamic weighted networks and reduce the time and space cost of the index.

\subsection{WCF-Index Construction} 
\label{subsec:construction}
The main idea of constructing \textit{WCF-Index} is to build the $\theta$-$tree_{k,t}$ for each graph instance $G_t$ for $k\in [1,k_{max}]$, where $k_{max}$ denotes the maximum core number of the vertex in $G_t$.
To obtain the $\theta$-$tree_{k,t}$, we group the vertices by their $\theta$-threshold value and add the vertex groups as tree nodes into the $\theta$-$tree$ according to the $\theta$-threshold and connectivity of the tree nodes, i.e., the vertex groups.

Algorithm~\ref{algo:construct} presents the procedure of building the \textit{WCF-Index} $\mathcal{I}$.
For each graph instance $G_t$, we build $\theta$-$tree_{k,t}$ for each available $k$ by first computing $\theta$-threshold of vertices and then construct and insert tree nodes to the  $\theta$-$tree$ index.
We first initialize two graphs $G_{pre}$ and $G_{cur}$ to store intermediate states of edge filtering (line 3).
Then we iteratively pick $\theta'\in\Theta$ in descending order.
For each $\theta'$, we get the edge set $E_{t,\theta'}$ whose weights are no less than $\theta'$ and obtain the induced  subgraph $G_t[E_{t,\theta'}]$ as the current state $G_{cur}$.
We can obtain a set of vertices $V_{\theta'}$ whose core number in $G_{cur}$ is increased with regards to the core number in the last state $G_{pre}$.
This implies that for a vertex $w\in V_{\theta'}$, the $\theta$-threshold of $w$ is $\theta'$ w.r.t. its increased core number (lines 4-7).
After that, we set the previous state $G_{pre}$ to be $G_{cur}$ and obtain the distinct values of the newly increased core numbers $K$ (lines 8-9).
Then, according to the newly identified $\theta$-threshold, we can construct and add tree nodes to the $\theta$-$tree_{k',t}$ for $k'\in K$ (lines 10-21).
To do that, we get $\theta$-$tree_{k',t}$ from $\mathcal{I}[k'][t]$, then we identify the groups of connected vertices whose $\theta$-threshold at $k'$ is $\theta'$ as the tree node $\mathbb{X}$ (lines 11-14).
To determine the position of $\mathbb{X}$, we find each tree node $\mathbb{Y}$ that contains any neighbor $v$ of $G_t[\mathbb{X}]$ and its root $\mathbb{Z}$, so that $\mathbb{X}, \mathbb{Y},\mathbb{Z}$ are connected (lines 15-18).
If $\mathbb{Z}.\theta>\mathbb{X}.\theta$, $\mathbb{X}$ is assigned as the parent of $\mathbb{Z}$, otherwise, their $\theta$-threshold are the same because the smaller $\theta'$ has not been visited yet, so we need to merge $\mathbb{X}$ to $\mathbb{Z}$ (lines 19-21).
After iterating all the standard threshold values, we can construct all the $\theta$-$tree_{k,t}$ completely for each possible $k$ of each $G_t$ and return the \textit{WCF-Index} $\mathcal{I}$.
The space cost of \textit{WCF-Index} is $O(\sum_{t_i\in T}\sum_{u\in V}core(u,G_{t_i}))$ as each vertex $u$ appears $core(u,G_{t_i})$ times in each graph instance.

\begin{algorithm}[ht]
\KwIn{A dynamic weighted graph $\mathcal{G}=\{G_{t_1}, G_{t_2},...G_{T}\}$}
\KwOut{The \textit{WCF-Index} $\mathcal{I}$}
$\mathcal{I}\gets\emptyset$\;
\For{$G_t\in\mathcal{G}$}{
$G_{pre}\gets\emptyset$; $G_{cur}\gets\emptyset$\;
\For{$\theta'\in\Theta$ in descending order}{
    $E_{t,\theta}\gets\{e\in E_{t}|W_{t}(e)\geq\theta\}$\;
    $G_{cur}\gets G_t[E_{t,\theta}]$\;
    $V_{\theta'}\gets\{u\in V|core(u,G_{cur})>core(u,G_{pre})$\}\;
    $G_{pre}\gets G_{cur}$\;
    $K\gets$ set of values $\{core(u,G_{cur})|u\in V_{\theta'}\}$\;
    \For{\textbf{each} $k'\in K$}{
        $\theta$-$tree_{k',t}\gets \mathcal{I}[k'][t]$\;
        $H\gets\{v\in V_{\theta'}|\theta$-$thres_{k'}(v,G_t)=\theta'\}$\;
        \For{\textbf{each} connected vertex set $\mathbb{X}\subseteq H$}{
        $\mathbb{X}$ as a new tree node of $\theta$-$tree_{k',t}$\;
            \For{\textbf{each} $v\in N(G_t[\mathbb{X}])$}{
                \If{$\theta$-$thres(v,G_t)>\theta'$}{
                    $\mathbb{Y}\gets$ get the node containing $v$\;
                    $\mathbb{Z}\gets$ get the root of $\mathbb{Y}$\;
                    \lIf{$\mathbb{Z}.\theta>\mathbb{X}.\theta$}{
                        $\mathbb{Z}.parent\gets\mathbb{X}$
                    }
                    \lElse{merge node $\mathbb{X}$ to $\mathbb{Z}$}
                }
            }
        }
        $\mathcal{I}[k'][t]\gets\theta$-$tree_{k',t}$\;
    }
}
}
 \textbf{Return} $\mathcal{I}$
\caption{$\theta$-Tree Construction\label{IR}}
\label{algo:construct}
\end{algorithm}

\begin{example}
Consider the graph instance $G_{t_1}$ in Figure~\ref{fig:intro}, we calculate the $\theta$-threshold of its vertices by inducing $G_{t_1}[E_{{t_1},\theta}]$ with increasing $\theta$.
Upon inducing $G_{t_1}[E_{{t_1},0.5}]$, comparing to $G_{t_1}[E_{{t_1},0.6}]$, we can observe the core number increase of $v_0$ and $v_1$ from 1 to 2, which implies $\theta$-$thres_2(v_0,G_{t_1})=\theta$-$thres_2(v_0,G_{t_1})=0.5$.
So we add tree node to $\theta$-$tree_2$ because the core number is increased to 2.
One node $\mathbb{X}=\{v_0,v_1\}$ is constructed as $v_0$ and $v_1$ are connected.
$\theta$-$tree_{2,{t_1}}$ has two nodes $\mathbb{Y}_1=\{v_2,v_3,v_4\}$ and $\mathbb{Y}_2=\{v_7,v_8,v_9\}$ from previous steps and the node that contains neighbors of $\mathbb{X}$ is $\mathbb{Y}_1$, whose root is itself.
$\mathbb{X}$ can thus be added as the parent of $\mathbb{Y}_1$ since $\mathbb{Y}_1.\theta>\mathbb{X}.\theta$.
By now, $\theta$-$tree_2$ contains three tree nodes.
After inducing $G_{t_1}[E_{{t_1},0}]$ we can construct $\theta$-$tree_{2,{t_1}}$ as Figure~\ref{fig:theta-tree}. $\theta$-$tree_{1,{t_1}}$ and $\theta$-$tree_{3,{t_1}}$ will also be obtained.
\end{example}

\subsection{WCF-Index Maintenance}
\label{subsec:maintenance}

In general, dynamic networks might have subtle changes, i.e., a small percentage of edges and vertices change or update, in two consecutive timestamps. It is time-consuming to simply compute the $\theta$-threshold for all vertices and re-construct the index. Therefore, in this section, it is highly desirable to develop an index maintenance strategy and update the index using the small number of changed edges and vertices only.

To explore the relationships of vertices with regards to different core numbers, we are motivated by the work in \cite{sariyuce2016incremental} that proposed an incremental core number update method for an evolving graph. It supports to locate a small set of vertices whose core number will be affected by using the below two concepts.
The other works \cite{zhang2017fast,liu2020incremental,hua2019faster} also follow the similar concepts of \textit{subcore} and \textit{purecore} in \cite{sariyuce2016incremental} to maintain the core numbers.

 \begin{definition}[\textit{subcore} in \cite{sariyuce2016incremental}]
 Given a graph $G$=$(V,E)$ and a vertex $u\in V$, the \textit{subcore} of $u$ denoted as $S_u$, is a set of vertices having the same core number as $u$ and connected with $u$ via a path, where each vertex on the path has the same core number as $u$.
 \end{definition}

 \begin{definition}[\textit{purecore} in \cite{sariyuce2016incremental}]
 Given a graph $G$=$(V,E)$ and a vertex $u\in V$, the \textit{purecore} of $u$ denoted as $P_u$, is a set of vertices where each vertex $w\in P_u$ satisfies:
 \begin{enumerate}
     \item Condition 1: the core number $core(w,G)$ of $w$ is equal to the core number $core(u,G)$ of $u$.  
     \item Condition 2: $w$ has a set $W$ of neighbors whose core numbers are no less than $core(w,G)$, and $|W|$ is larger than $core(u,G)$. 
     \item Condition 3: $w$ is connected to $u$ via a path, where each vertex on the path satisfies the conditions (1) and (2).
 \end{enumerate}
 \end{definition}

Specifically, given two vertices $u$ and $v$ in a graph $G$=$(V,E)$, and $core(u,G)\leq core(v,G)$, if an edge $(u,v)$ is removed from $G$, then only the vertices in the subcore set $S_u$ may have their core number decreased; if an edge $(u,v)$ is added to $G$, then only the vertices in the purecore set $P_u$ may have their core number increased. Thus, we can extend the rules to the weighted graph, in which the updates include edge insertion, edge deletion, and edge weight change.

Considering that $G_{t_n'}$ is obtained by inserting an edge $(u,v)$ with weight $\theta'$ to $G_{t_n}=(V,E_{t_n},W_{t_n})$.
For a vertex $w\in V$, if $\theta$-$thres_k(w,G_{t_n})=\theta''$ and $\theta''\geq\theta'$, then $\theta$-$thres_k(w,G_{t_n'})$ will remain unchanged.
\begin{property}[Insertion of an Edge]
\label{prop:insert}
Given a graph $G_{t_n}$ and $\theta$-$tree_{k,t_n}$ for each $k\in[1,k_{max}]$, and two vertices $u$ and $v$ such that $\theta$-$thres_k(u,G_{t_n})\leq\theta$-$thres_k(v,G_{t_n})$, if an edge $(u,v)$ is inserted with weight $\theta'$, then only the vertices $\{w\in P_u| \theta$-$thres_k(w,G_{t_n})<\theta'\}$ may have their $\theta$-threshold increased. 
\end{property}

\begin{property}[Deletion of an Edge]
Given a graph $G_{t_n}$ and $\theta$-$tree_{k,t_n}$ for each $k\in[1,k_{max}]$, and two vertices $u$ and $v$ such that $\theta$-$thres_k(u,G_{t_n})\leq\theta$-$thres_k(v,G_{t_n})$, if an edge $(u,v)$ is removed with weight $\theta'$, then only the vertices $\{w\in S_u| \theta$-$thres_k(w,G_{t_n})\leq\theta'\}$ may have their $\theta$-threshold decreased. 
\end{property}

The edge insertion and edge deletion can be treated as the update of edge weight. 

\begin{property}[Update of Edge Weight]
Given a graph $G_{t_n}$ and $\theta$-$tree_{k,t_n}$ for each $k\in[1,k_{max}]$, two vertices $u$ and $v$ such that $\theta$-$thres_k(u,G_{t_n})\leq\theta$-$thres_k(v,G_{t_n})$, we have (1) if the weight of edge $(u,v)$ increases from $\theta_1$ to $\theta_2$ (w.r.t. $\theta_1<\theta_2$), then only the vertices $\{w\in P_u| \theta$-$thres_k(w,G_{t_n})\in[\theta_1,\theta_2)\}$ may have their $\theta$-threshold increased;
(2) if the weight of edge $(u,v)$ decreases from $\theta_2$ to $\theta_1$ (w.r.t. $\theta_1<\theta_2$), then only the vertices $\{w\in S_u| \theta$-$thres_k(w,G_{t_n})\in(\theta_1,\theta_2]\}$ may have their $\theta$-threshold decreased. 
\end{property}

\begin{figure}[tb]
  \centering
  \includegraphics[width=0.5\textwidth]{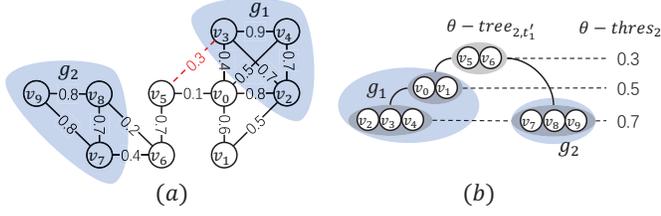}
  \caption{$\theta$-tree of $G_{t_1'}$, $k$=2}
  \label{fig:theta-tree2}
\end{figure}

\begin{example}
Figure~\ref{fig:theta-tree2}(a) shows an updated graph instance of $G_{t_1'}$ by adding an edge $(v_3, v_5)$ with weight 0.3 to $G_{t_1}$.
We can identify $P_{v_3}=\{v_0,v_2,v_3,v_5,v_6,v_7,v_8\}$ in $G_{t_1}$. In addition, the $\theta$-threshold of $v_5$ or $v_6$ is less than 0.3 and other vertices' $\theta$-threshold is no less than 0.3. According to Property~\ref{prop:insert}, only $v_5$ and $v_6$ may have their $\theta$-threshold increased.
After recalculating $\theta$-$thres_2(v_5,G_{t_1'})$ and $\theta$-$thres_2(v_6,G_{t_1'})$, we can update the tree index from Figure~\ref{fig:theta-tree} to Figure~\ref{fig:theta-tree2}(b).
\end{example}

\subsection{WCF-Index Compression}
\label{subsec:compression}
Sometimes, the graph instances of some consecutive timestamps may be similar because the edge weight and graph structure change progressively over time.
Besides that, one tree node usually contains multiple vertices as it gathers many connected vertices with the same threshold. It is likely to have much duplicate information across $\theta$-$tree$ indices. Thus, we need to develop an index compression strategy in order to reduce the redundancy.
The key idea is to utilize a virtual node to replace the tree node that contains multiple vertices and appears frequently. The actual vertices of the virtual nodes are stored in an auxiliary table.

\begin{figure}[tb]
  \centering
  \includegraphics[width=0.45\textwidth]{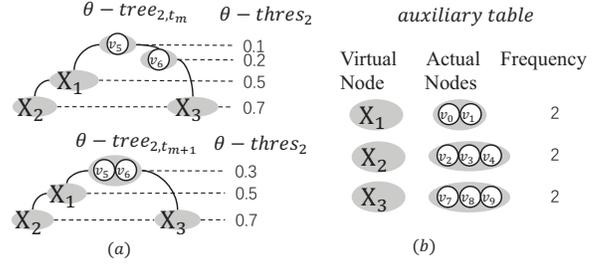}
  \caption{Compressed \textit{WCF-Index} and auxiliary table}
  \label{fig:auxiliary}
\end{figure} 

\begin{example}
Figure~\ref{fig:auxiliary} shows an example of compressed $\theta$-$tree_{2,t_1}$ and $\theta$-$tree_{2,t_1'}$ where the nodes with vertices $\{v_0,v_1\}$, $\{v_2,v_3,v_4\}$, $\{v_7,v_8,v_9\}$ are replaced by virtual nodes $\mathbf{X_1}$, $\mathbf{X_2}$, $\mathbf{X_3}$.
\end{example}

Each virtual node can be regarded as an encoding of a unique vertex set.
The space cost can be reduced if we use an auxiliary table to only maintain one copy of the tree nodes that frequently appear in \textit{WCF-Index}, and keep a virtual id at the positions of these nodes in \textit{WCF-Index}.
To make the compression, we only select the tree nodes that can bring positive space gain.
\begin{definition}[Space Gain]
The space gain is defined as the space saved from replacing the tree node $\mathbb{X}$ as a virtual node $\mathbf{X}$, that is 
$$SG(\mathbb{X}) = f*(|\mathbb{X}|-1)-|\mathbb{X}| $$
where $f$ is the frequency of $\mathbb{X}$ that appears in the \textit{WCF-Index}, and $|\mathbb{X}|$ is the size the vertex set.
\end{definition}

By scanning and counting the frequency of tree nodes, we can calculate the space gains for all the tree nodes, and generate the compressed index easily. We do not provide the pseud codes in this paper due to the limited space.

%% file: Experiment.tex
\newcommand{\data}{\textcolor{cyan}{[DATA]}}

\section{Experiment}
\label{sec:exp}

We conduct extensive experiments to evaluate the performance of our proposed algorithms, including EEF-based Online CRC Search in Algorithm~\ref{algo:edge}, \textit{WCF-Index} based CRC Search in Algorithm~\ref{algo:index_query} and $\theta$-Tree Construction in Algorithm~\ref{algo:construct}, denoted as \textit{EEF-CRC}, \textit{WCF-CRC} and \textit{WCF-Construct}, respectively.
We implement a baseline method based on maximal spaning core (\textit{SpanCore}) \cite{galimberti2018mining}, which calculates all the k-core subgraphs with different $k$ value in various time intervals. We additionally remove the edges and calculate the reliability score of each candidate subgraph to get the optimal results.
We compare the effectiveness of our proposed community model with PC \cite{li2018persistent} and SC \cite{9001192}.
We also evaluate the effectiveness of index maintenance (\textit{WCF-Maintain}) and index compression.
All the experiments are conducted on a Windows machine with an Intel i9-10900F CPU @ 2.80GHz and 32.0 GB DDR4-RAM.

\subsection{Experimental Setup}\label{sec:expSetup}

\begin{table*}[tb]
\caption{Dataset Statistics}
\center
\renewcommand\arraystretch{1}
\scalebox{0.9}{
\begin{tabular}{c|c|c|c|c|c|c|c|c}
\hline
& BitcoinAlpha & BitcoinOtc  & Retweet & TAT    & Email & Reddit & HepPh   & StackOverFlow \\ \hline
$|V|$ / $\widehat{|V|}$  & 3,783 / 688  & 5,881 / 997 & 18,470 / 4,249 & 34,761 / 3,613  & 986 / 658 & 55,863 / 4,908  & 28,093 / 4,009   & 2,601,977 / 79,629 \\ \hline
$|E|$ / $\widehat{|E|}$  & 24,186 / 1,497 & 35,592 / 2,247     & 61,157 / 5,554 & 171,403 / 5,122 & 332,334 / 2,619 & 571,927 / 8,556 & 4,596,803 / 78,535 & 63,497,050 /325,080 \\ \hline
$|T|$       & 10             & 10                   & 30      & 30     & 30        & 30     & 50      & 100  \\ \hline
$\widehat{density}$ & 0.0065    &   0.0047      & 0.0006   & 0.0015   & 0.0121   & 0.0007   & 0.0099  & 0.0002 \\ \hline
$\widehat{k}_{max}$ & 6.6            & 8.5                 & 4.4     & 12.6   & 7.9       & 12.4   & 118.2   & 27.1 \\ \hline
$\widehat{k}_{query}$ &    3.5   &   4.3    &   1.9   &  7.1  &   3.9    &  5.5  & 59.4  &  10.2   \\ \hline
\end{tabular}
}
\label{tab:data}
\end{table*}

\textbf{Datasets.} We conduct the experiments on eight real-world dynamic network datasets collected from SNAP\footnote{https://snap.stanford.edu/data/} and Network Data Repository\footnote{https://networkrepository.com}. 
In \textit{BitcoinAlpha (BA)} and \textit{BitcoinOTC (BO)} datasets, the edge weight represents the rating between two users. 
In the remaining datasets, the edge weight is calculated from the interaction frequency.
The edge weight of all the datasets is normalized to $[0,1]$ by min-max normalization.
The statistics of the dataset are shown in Table~\ref{tab:data}.
The number of vertices and edges are denoted as $|V|$ and $|E|$, respectively.
For each dataset, we first sort the edges by chronological order, and then divide them into $|T|$ partitions, i.e., $|E|/|T|$ edges, where $|T|$ is the target number of graph instances. It guarantees each graph instance contains meaningful $k$-$core$ components.
For example, the largest dataset \textit{StackOverFlow (SOF)} is divided into $|T|=100$ snapshots and the medium-sized datasets, e.g., \textit{TechAsTopology (TAT)}, \textit{Retweet}, etc. are divided into $|T|=30$ instances. 
We denote $\widehat{|V|}$, $\widehat{|E|}$, $\widehat{density}$ and $\widehat{k}_{max}$ as the average of vertex numbers, edge numbers, density, and the largest core numbers of the $|T|$ graph instances, respectively. 

\begin{table}[tb]
\caption{Parameters and default values}
\renewcommand\arraystretch{1}
\center
\scalebox{0.9}{
\begin{tabular}{c|c|c}
\hline
    Parameter & Values & Description \\ \hline
    $k$ & 20\%, \textbf{40\%}, 60\%, 80\% & \% of $k_{max}$\\ \hline
    $\theta$ & 0.0,0.2,\textbf{0.4},0.6,0.8 & weight threshold\\ \hline
    $t$ & 4, 8, \textbf{12}, 16, 20 & Time span (snapshots)\\ \hline
\end{tabular}}
\label{tab:param}
\end{table}

\textbf{Parameters.} 
Table~\ref{tab:param} shows the detailed setting of the parameters used in the experiments. 
To better fit the dataset and cover more meaningful situations, we vary the query parameter $k$ as $20\%$, $40\%$, $60\%$, $80\%$ of $\widehat{k}_{max}$ for each dataset with the default value $40\%$.
The threshold value varies from 0.0 to 0.8 with the default value of 0.4. The length of the query time interval was specified as 4, 8, 12, 16, 20 with the default value of 12.
Their default values are marked in bold font. 
We also vary the $\alpha$ parameter from 0 to 6 to show its effect on the returned community.
We sample 100 query vertices whose core numbers are uniformly distributed in $[1,k_{max}]$ for each dataset and report their average running time as the time cost.
The average core number of the 100 query vertices is shown in Table~\ref{tab:data} as $\widehat{k}_{query}$.
In general, $\widehat{k}_{query}$ is around 50\% of $k_{max}$, which reflects the common scenario of query vertex. 

\subsection{Evaluation of Query Efficiency}
\label{subsec:efficiency}
\begin{figure}[tb]
     \centering
     \includegraphics[width=0.4\textwidth]{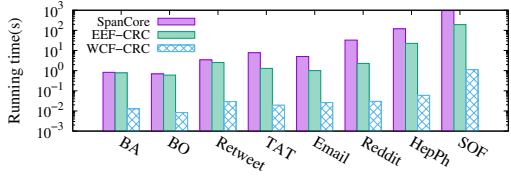}
     \caption{CRC Query Time on All Datasets}
     \label{fig:overall}
\end{figure}

In this section, we present the performance of SpanCore, EEF-CRC, and WCF-CRC under the default parameter settings. Figure~\ref{fig:overall} demonstrates the time cost when we run the three algorithms over eight datasets. 
Both of the proposed algorithms outperform the baseline algorithm \textit{SpanCore}.
\textit{EEF-CRC} is slightly faster than \textit{SpanCore}, as they both need to determine the core number of the vertices but \textit{EEF-CRC} only requires the local information of the query vertex.
\textit{WCF-CRC} runs much faster than \textit{EEF-CRC}. For instance, \textit{WCF-CRC} reduces the time cost of \textit{EEF-CRC} by about 89 times. 

To show the impact of each parameter, we also evaluate the efficiency of the proposed algorithms by varying the values of parameters $k$, $\theta$, and $t$, respectively.
We utilize two representative datasets \textit{Reddit} and \textit{SOF} to demonstrate the experimental results.

\begin{figure}[tb]
     \centering
         \includegraphics[width=.45\textwidth]{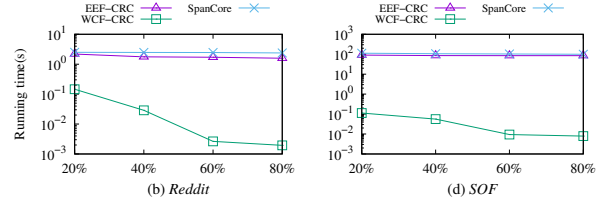}
     \caption{CRC Query Time with Varying $k$}
     \label{fig:vary_k}
\end{figure}

\underline{Varying $k$.} Figure~\ref{fig:vary_k} shows the average time cost of our proposed algorithms when $k$ varies from $20\%$ to $80\%$ of the corresponding $\widehat{k}_{max}$ values. 
\textit{WCF-CRC} is significantly more efficient than the other two algorithms, and \textit{SpanCore} consumes the most time in all settings.
For instance, \textit{SpanCore} takes 2.46s, \textit{EEF-CRC} takes 1.76s, while \textit{WCF-CRC} only needs 0.03s to complete the query processing in \textit{Reddit} dataset where $k$ is 40\% of the average large core number (i.e., $k=5$).
With the increase of $k$, all the algorithms consume decreasing time.
But \textit{SpanCore} and \textit{EEF-CRC} are less sensitive to $k$ than \textit{WCF-CRC} because they need to scan all the edges and compute the core numbers of the vertices induced by the edges, while \textit{WCF-CRC} can directly retrieve the core numbers using \textit{WCF-Index}. 

\begin{figure}[tb]
     \centering
         \includegraphics[width=.45\textwidth]{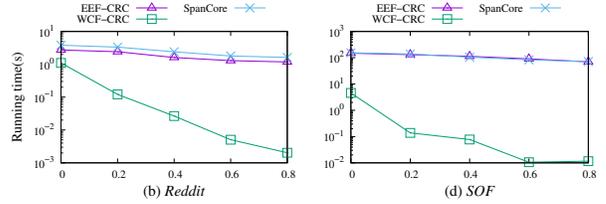}
     \caption{CRC Query Time with Varying $\theta$}
     \label{fig:vary_theta}
\end{figure}

\underline{Varying threshold $\theta$.}
Figure~\ref{fig:vary_theta} shows the average time cost of our proposed algorithms when the threshold $\theta$ varies from $0$ to $0.8$. 
\textit{WCF-CRC} outperforms \textit{SpanCore} and \textit{EEF-CRC} significantly.
For instance, in \textit{Reddit}, when $\theta=0.2$, \textit{SpanCore} takes 3.33s, \textit{EEF-CRC} takes 2.41s while \textit{WCF-CRC} takes 0.12s. \textit{WCF-CRC} is faster than the other two by more than 20 times. 
When $\theta$ is given as 0.8, the three algorithms take 1.62s, 1.17s and 0.002s, respectively, i.e., 
\textit{WCF-CRC} can reduce the time cost of \textit{EEF-CRC} by about 500 times.
With the increase of $\theta$, the speedup trend of \textit{WCF-CRC} becomes significant because there are small number of tree nodes in \textit{WCF-Index}  when $\theta$ is set as a large value, i.e., more vertices can be pruned.
Similarly, \textit{SpanCore} and \textit{EEF-CRC} also consume less time because the significant number edges can be pruned with the higher $theta$ threshold. However, scanning all the edges in the locally connected subgraph for each query vertex is inevitable.

\begin{figure}[tb]
     \centering
         \includegraphics[width=.45\textwidth]{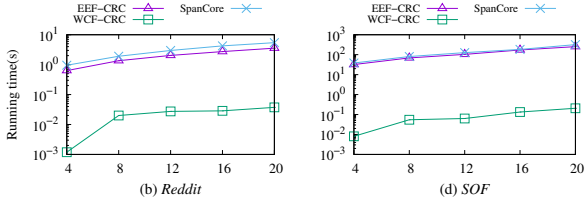}
     \caption{CRC Query Time with Varying $|T_Q|$}
     \label{fig:vary_t}
\end{figure}

\underline{Varying time span $|T_Q|$.} 
Figure~\ref{fig:vary_t} shows the average time cost of the proposed algorithms when the time span $|T_Q|$ varies from 4 to 20. 
All the algorithms consume higher time cost when the query time span $|T_Q|$ increases.
For instance, in \textit{Reddit} dataset, \textit{SpanCore} takes 0.96s, 1.91s, 2.98s, 4.24s, and 5.44s. \textit{EEF-CRC} takes 0.63s, 1.35s, 2.04s, 2.73s and 3.55s, respectively. But, \textit{WCF-CRC} only takes 0.0012s, 0.02s, 0.027s, 0.029s and 0.037s at the same settings.
The growth of runtime is consistent with the time complexity in Section~\ref{Sec:Online} and Section~\ref{subsec:dyna}.

\begin{figure}[tb]
     \centering
     \includegraphics[width=0.45\textwidth]{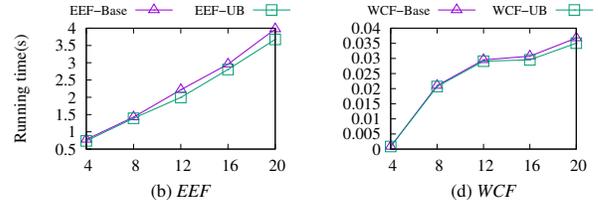}
     \caption{Upper Bound Evaluation with Varying $|T_Q|$}
     \label{fig:ub_eff}
   \end{figure}

\underline{Efficiency of Upper Bound.} 
Figure~\ref{fig:ub_eff} shows the efficiency of the proposed algorithms with (i.e., \textit{EEF-UB}, \textit{WCF-UB}) and without (i.e., \textit{EEF-Base}, \textit{WCF-Base}) using the upper bound pruning strategy on the \textit{Reddit} dataset with varying $|T_Q|$.
We can find that the pruning capability of the upper bound can be accelerated with the increase of the query time interval. 
For instance, \textit{EEF-UB} is faster than \textit{EEF-Base} by 5\% when $|T_Q|=4$, but the acceleration can achieve by 10\% when $|T_Q|=20$.  Compared to \textit{WCF-Base}, the improvement of \textit{WCF-UB} is not significant.
Since the nature of the upper bound is to estimate the maximum reliability score of the community candidates for each potential time interval,  
there are more chances to prune more intermediate community candidates when query time interval is large.

\subsection{Evaluation of Index Construction and Maintenance}
\begin{figure}[tb]
\begin{minipage}[t]{0.49\linewidth}
    \includegraphics[width=\linewidth]{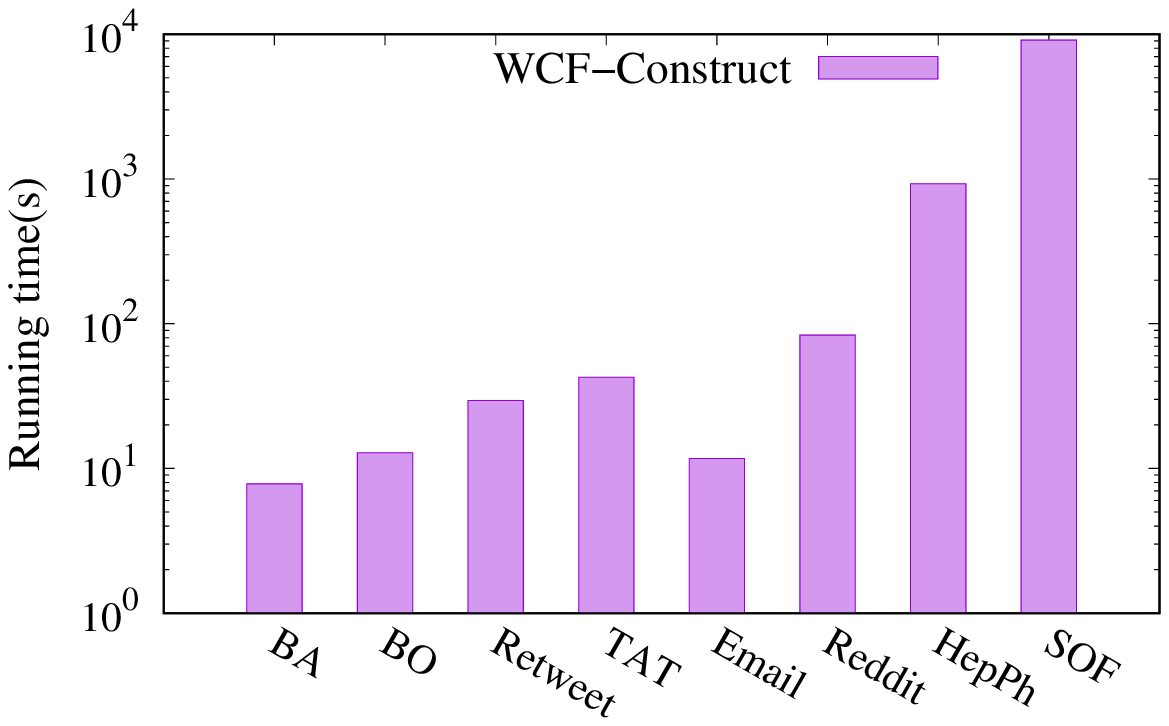}
    \caption{Index Construction Time on All Datasets}
     \label{fig:const}
\end{minipage}
\begin{minipage}[t]{0.49\linewidth}
    \includegraphics[width=\linewidth]{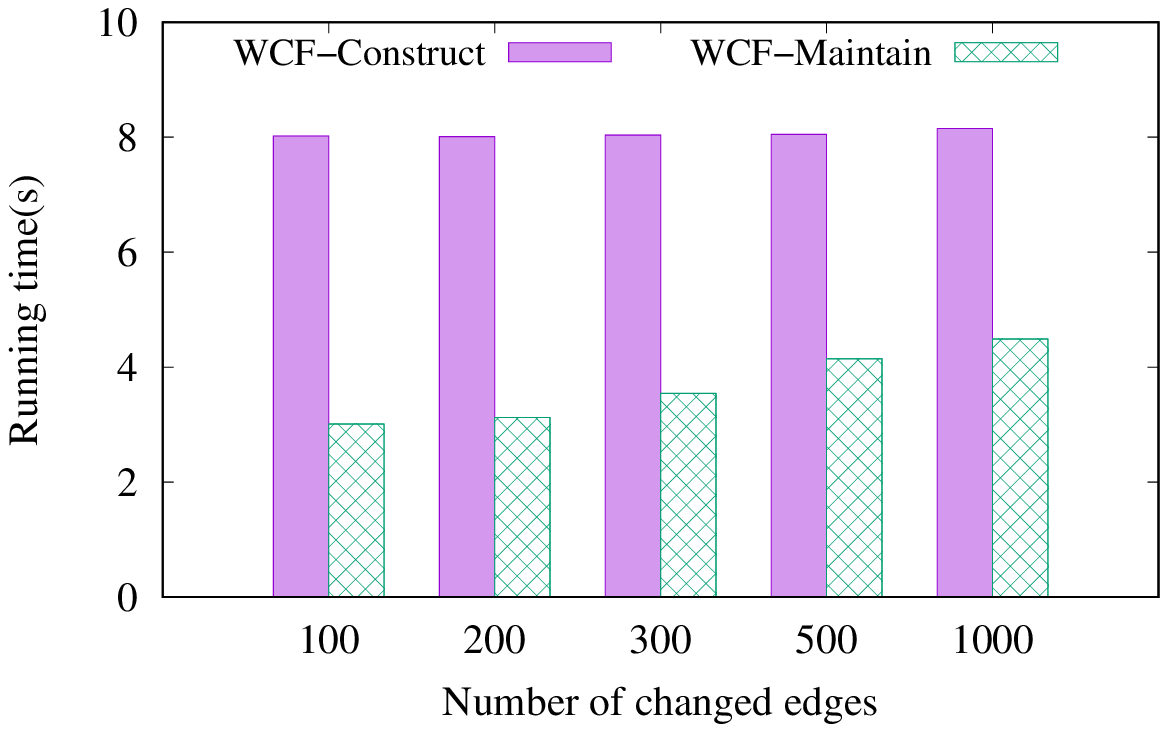}
    \caption{Index Construction Time by Maintenance}
    \label{fig:main}
\end{minipage} 
\end{figure}

In Figure~\ref{fig:const}, we report the  runtime of \textit{WCF-Construct} for all the graph instances of eight datasets. Generally, it takes around 10s for datasets with small number of vertices like \textit{BA}, \textit{BO}, and \textit{Email}. For the largest dataset like \textit{SOF}, it needs 25h to complete the index construction.

We also evaluate the effectiveness of the index maintenance method.
Taking the first graph instance of \textit{Reddit} as the base, we randomly sample 100, 200, 300, 500, 1000 edges and mix the operation of edge insertion, deletion, and weight update to generate a synthetic instance.
Figure~\ref{fig:main} shows the time cost of \textit{WCF-Construct} and \textit{WCF-Maintain} on the synthetic graph, where the speed up of \textit{WCF-Maintain} is significant.
For instance, reconstructing the index takes 8s, but \textit{WCF-Maintain} only takes 4s when 1000 edges are updated.

\subsection{Evaluation of Scalability}
\begin{figure}[tb]
     \centering
     \includegraphics[width=0.4\textwidth]{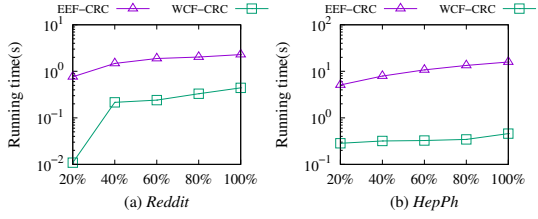}
     \caption{CRC Query Time with Different Sampling Ratios}
     \label{fig:sca1}
\end{figure}

We evaluate the scalability of proposed algorithms including \textit{EEF-CRC}, \textit{WCF-CRC} and \textit{WCF-Construct} by using five graph instances from two datasets \textit{Reddit} and \textit{HepPh}. 
For each dataset, we generate four new datasets with different sizes by randomly sampling 20\%, 40\%, 60\%, 80\% edges from the dataset, respectively. The dataset itself is considered with the 100\% data size. 
Figure~\ref{fig:sca1} shows the time cost of \textit{WCF-CRC} and \textit{EEF-CRC} on the size-varying datasets.
With the increase of the data size, we can find that the running time of \textit{WCF-CRC} and \textit{EEF-CRC} grow in a gentle trend, which implies that both algorithms are easily applicable to large-scale networks.

\begin{figure}[tb]
     \centering
     \includegraphics[width=0.4\textwidth]{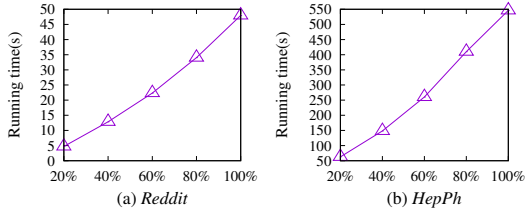}
     \caption{Index Construction Time with Different Sampling Ratios}
     \label{fig:sca2}
\end{figure}

Furthermore, we also show the scalability of index construction in Figure~\ref{fig:sca2}. From this, we can observe a linear increasing trend of the construction time.
For instance, the index constructing time of \textit{Reddit} is 4.7s, 12.8s, 22.3s, 34.0s and 47.9s when the size of the dataset increases as 20\%, 40\%, 60\%, 80\% and 100\%, respectively.

\begin{table*}[tb]
\caption{\label{tab:index_size}Index Size \& Compression (kb)}
\renewcommand\arraystretch{1}
\centering
\scalebox{0.9}{
\begin{tabular}{c|c|c|c|c|c|c|c|c}
\hline
 & BitcoinAlpha & BitcoinOtc  & Retweet & TAT    & Email & Reddit & HepPh   & StackOverFlow \\ \hline
Raw Data   &  1,108     & 1,648   &  4,935  & 13,619 &  5,324  &  37,376   &  263,168      &  1,153,024        \\ \hline
Original Index   &  333     & 645   &  1,569  & 7,047 &   1,139  &   45,216   &   252,561      &  874,931         \\ \hline
Compressed &  328       & 616   &  1,418  & 6,205 &  1,133   &  39,924    &   158,734      &   835,721        \\ \hline
Auxiliary Table &   3        & 11    &  120    & 430  &    3  &   834   &    11,406      &  5,361           \\ \hline
\end{tabular}
}
\end{table*}

\subsection{Evaluation of Index Size with Compression}
\label{subsec:size_compress}
We show the \textit{WCF-Index} size of the eight datasets in Table~\ref{tab:index_size}.
For each dataset, we take ten graph instances.
The largest dataset \textit{StackOverFlow} takes 874,931kb and the smallest dataset \textit{BitcoinAlpha} takes 333kb.
The compressed size is the sum of the compressed index and the auxiliary table.
A significant compression effectiveness can be observed in Table~\ref{tab:index_size}, e.g., the index size of \textit{HepPh} can be compressed to 50\% of the original size.

\subsection{Evaluation of Query Effectiveness}
\label{subsec:exp_effect}

\begin{table}[t]
\caption{\label{tab:case} Case Study on \textit{Reddit}}
\centering
\renewcommand\arraystretch{1}
\scalebox{0.9}{
\begin{tabular}{c|c|c|c|c|c}
\hline
t                  & \multicolumn{2}{c|}{Community (ASS)} & ASD          & ASCore        & ASCon         \\ \hline
\multirow{3}{*}{3} & \multicolumn{1}{c}{SC} & (65.4)  & 0.19         &  7.65         & 0.80          \\ 
                   & \multicolumn{1}{c}{PC} & (50.2)  & 0.17         &  5.5          & 0.85          \\ 
                   & \multicolumn{1}{c}{CRC} & (56) & \textbf{0.31}& \textbf{10.6} & \textbf{0.77} \\ \hline
\multirow{3}{*}{4} & \multicolumn{1}{c}{SC} & (30.4)  & 0.31         &  6.5          & 0.89          \\ 
                   & \multicolumn{1}{c}{PC} & (49.8)  & 0.23         &  7.3          & 0.83          \\ 
                   & \multicolumn{1}{c}{CRC} & (41) & \textbf{0.39}& \textbf{10.4} & \textbf{0.83} \\ \hline
\multirow{3}{*}{5} & \multicolumn{1}{c}{SC} & (23.2) & 0.40         &  6.8          & 0.91          \\ 
                   & \multicolumn{1}{c}{PC} & (48) & 0.26         &  7.9          &  0.84 \\ 
                   & \multicolumn{1}{c}{CRC} & (36) & \textbf{0.43}& \textbf{10.3} &   \textbf{0.83}        \\ \hline
\end{tabular}}
\end{table}

To show the effectiveness of finding communities in dynamic or temporal networks, we compare our ($\theta,k$)-$core$ reliable community (CRC) with the Persistent Community (PC) and Stable Community (SC) proposed by Li \cite{li2018persistent} and Qin \cite{9001192}, respectively. 
To do this, we select five graph instances of \textit{Reddit} dataset and return the largest community $C$ obtained by SC, PC, and our CRC.
To show the quality of returned communities, we utilize three community quality metrics:
\begin{itemize}[leftmargin=*]
    \item \textbf{Average Snapshot Density (ASD)} measures how dense is the community and captures the intuition that a good community should be closely connected inside. The larger is the density, the closer the community is connected. 
    Average snapshot density is calculated as the average density of the community in each snapshot: $ASD = \sum_{i=1}^t density(G_{t_i}[C])/t$.
    \item \textbf{Average Snapshot Core (ASCore)} captures the degree information of vertices and evaluates the closeness of the community. The larger is the core number, the more interactions each vertex will keep with others in the community.
    ASCore calculates the average value of the average core number of each vertex in each snapshot: $ASCore = \sum_{i=1}^t(\sum_{v\in V}core(v,G_{t_i}[C])/|V|)/t$.
    \item \textbf{Average Snapshot Conductance (ASCond)} measures how ``well-knit'' the graph is. The higher is the conductance, the easier the community can communicate with the vertices outside the community. In the local community detection task, the smaller conductance is desired as it implies the community is tightly self-capsulated.
    Here, the average snapshot conductance is calculated as the average conductance of the community in each snapshot: $ASD = \sum_{i=1}^t conductance(G_{t_i}[C])/t$.
\end{itemize}

Table~\ref{tab:case} shows the experimental results of evaluating the community quality on the \textit{Reddit} dataset. 
The community is obtained with the same structural cohesiveness constraint (core number or number of neighbors equals 8). 
We vary the duration or frequency of the community ($\tau$ in SC and PC, $d$ in CRC), denoted by $t$ in Table~\ref{tab:case}, to compare the community quality with different temporal features.
The size of each community is also provided as the average snapshot size \textit{ASS} in Table~\ref{tab:case}.
It can be observed that CRC performs best in all the measurements. 
For example, when $t=4$, CRC finds a community with the highest \textit{ASD} of 0.39, the highest \textit{ASCore} of 10.4 and the lowest \textit{ASCond} of 0.83.
When $t=4$ and $5$, PC and CRC have similar \textit{ASCond} score.
\textit{PC} generally finds the largest community at the cost of lower density and cohesiveness.
\textit{CRC} outperforms \textit{SC} with the larger community size and the closer connection.

\begin{figure}[tb]
    \centering
    \includegraphics[width=0.4\textwidth]{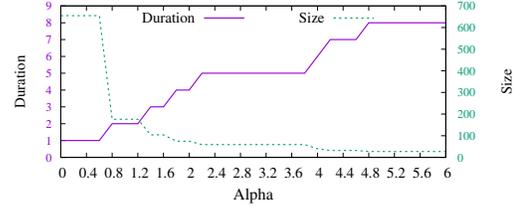}
    \caption{Community Size and Duration with Varying $\alpha$}
    \label{fig:alpha}
\end{figure}

\begin{figure}[tb]
  \centering
  \includegraphics[width=0.4\textwidth]{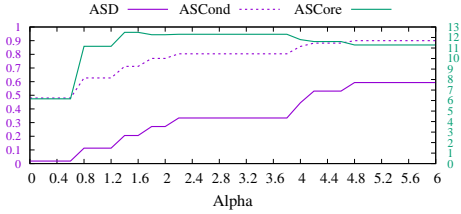}
  \caption{Community Quality with Varying $\alpha$}
  \label{fig:alpha_quality}
\end{figure}

\underline{Community with varying $\alpha$}.
Figure~\ref{fig:alpha} shows an example of obtained reliable community by querying the vertex \textit{funny} in the \textit{Reddit} dataset where $\alpha$ varies from 0 to 6.
With the increase of $\alpha$, the duration of the optimal CRC increases and the community size decreases.
The progressive change of community duration shows that parameter $\alpha$ is able to smoothly adjust the balance between community size and duration.
Figure~\ref{fig:alpha_quality} shows the trending change of the selected three quality metrics $ASD, ASCore$ and $ASCond$ when $\alpha$ increases. All the scores increase significant when $\alpha$ varies from 0 to 2. After that, their trends become steady relatively.

%% file: Related_Works.tex
\section{Related Works}
\label{Sec:RelatedWork}

\textbf{Local Community Search in Static and Time-varying Networks.}
Local community search has been studied in many existing works.
In static networks, existing methods can be classified into two categories.
The first method is based on random walk, which aims to assign scores to the vertex from the query vertex and identify the local community based on the scores.
Wu et al. \cite{wu2015robust} used a single random walker,
and Bian et al. \cite{bian2017many} introduced multiple walkers to assign vertex scores based on the hitting probability.
Bian \cite{bian2018multi} further proposed memory based multiple walker that records the entire visiting history and supports multiple local communities w.r.t. different query vertices simultaneously.
Another method is based on capturing cohesiveness structures \cite{huang2017community} such as \textit{k-truss} \cite{huang2014querying,akbas2017truss, liu2020truss}, \textit{k-core} \cite{cui2014local,barbieri2015efficient,li2015influential} and \textit{k-clique} \cite{cui2013online,shan2016searching}. 
To deal with the changes of network data over time, Takaffoli et al. \cite{takaffoli2013incremental} explored local community mining in the dynamic social network by extending \textit{L-metric} \cite{chen2009detecting} to an incremental version. 
Luo et al. \cite{luo2018local} divided the formation of the local community into three stages and designed different dynamical membership functions to construct the local community with better cohesiveness.
DiTursi et al.~\cite{ditursi2017local} proposed PHASR method to detect local community in the dynamic networks, and Papadopoulos \cite{n2020distributed} expanded PHASR to fit the distributed processing standard of Apache Spark engine.

\textbf{K-core Community Search.}
In this work, we consider the community size (number of vertices) and duration (continuity of the vertex engagement) as two important factors, so we use \textit{k-core} model which is defined on vertex attributes rather than other classic models like \textit{k-truss} that defined on edge attributes.
\textit{K-core} was firstly introduced by Seidman et al. \cite{seidman1983network} and becomes one of the most widely used measurements of graph cohesiveness.
Seideman et al. \cite{sozio2010community} developed a greedy algorithm to discover the dense subgraph by iteratively removing the vertices with the minimum degrees.
Batagelj et al. \cite{batagelj2003m} proposed a linear core decomposition algorithm to compute the core number of all the vertices. 
Cui et al. \cite{cui2014local} developed a local community search algorithm that starts from a query vertex $q$ and spans iteratively to include the local optimal vertex into the community.
Barbieri et al \cite{barbieri2015efficient} proposed an index structure based on the nested feature of the core number and improved the community search significantly. Following \cite{barbieri2015efficient}, Fang et al. \cite{fang2016effective} improved the efficiency of index construction. 
To include network dynamics, Li et al. \cite{li2013efficient} devised a core maintenance algorithm in large dynamic networks.
Wu \cite{wu2015core} proposed distributed algorithms based on the block-centric model to compute cores in the temporal graph. 
Galimberti et al. \cite{galimberti2018mining} identified all the maximal $k$-core with various time span and k value. Based on \cite{galimberti2018mining}, Hung and Tseng \cite{hung2021maximum} extended to the maximal lasting k-core. However, these works mainly focused on core decomposition in dynamic graphs and didn't consider the edge weight and query vertex like this work.

\textbf{Cohesive Subgraph Mining in Dynamic Networks.} Our work also relates to dense subgraph mining, which aims to identify the densely connected vertices in temporal or dynamic networks.
Abdelhamid et al. \cite{abdelhamid2017incremental} proposed an incremental approach called IncGM+ to extend the traditional Frequent Subgraph Mining (FSM) method into dynamic networks by only updating the ``fringe'' patterns.
Ma et al. \cite{ma2017fast} proposed a fast computation algorithm to identify dense subgraphs in temporal graphs where edges have positive or negative weights.
However, their method relies on the ``evolving convergence phenomenon'' that assumes weights of all edges are increasing or decreasing in the same direction which is too strict for the real world. And they didn't consider the community continuity.
Semertzidis and Pitoura \cite{semertzidis2018top} proposed the problem of querying the frequent subgraph patterns in the directed dynamic networks and returns the top-$k$ durable matches.
Liu et al. \cite{liu2019finding} considered the duration of the found dense subgraphs using an expectation-maximization method. They are unable to deal with edge weight, which is limited in real applications. 

%% file: Conclusion.tex
\section{Conclusions and Future Work}
\label{Sec:Conclusion}

In this paper, we first discussed the necessity of reliable local community in dynamic networks and proposed the novel most reliable community search problem. Then, we developed an online $(\theta,k)$-$core$ reliable community search approach by pruning the ineligible edges based on the given threshold and their lasting times. After that, we designed an effective \textit{WCF-Index} to maintain the vertex candidates of $(\theta,k)$-$core$ subgraphs, and developed an efficient index-based dynamic programming approach. Finally, the empirical evaluations on a variety of datasets and parameter settings illustrate the efficiency and effectiveness of the proposed approaches.
In this work, we mainly focus on single quey vertex situation. 
However, our proposed algorithms can be extended to support querying a set of vertices.
For \textit{EEF} algorithm, we can start the edge search from all query nodes simultaneously, and maintain a visited edge set to avoid repeat traverse.
Then we can follow the EEF algorithm to return the valid local k-core.
For \textit{WCF-Index} query algorithm, multiple query nodes can be easily supported by filtering the intermediate result that does not contain all the query nodes.
To explore the significance of multiple query nodes in community discovery, one potential research direction is to investigate the local engagement of the query nodes and identify the meaningful communities.